\documentclass[
aps,
prb,
twocolumn,
superscriptaddress,%
showpacs,%
preprintnumbers,%
byrevtex,%
floatfix%
]{revtex4-2}

\usepackage{dcolumn}
\usepackage{bm}
\usepackage{ifthen}
\usepackage[usenames]{color}
\usepackage{amssymb}
\usepackage{textcomp}
\usepackage{amsfonts}
\usepackage{amsmath}
\usepackage{graphicx}
\usepackage{graphpap}
\usepackage{placeins}  

\begin{document}
	
\date{\today}
\title{Theoretical limits of magnetic detection of structural surface defects at the nanometer scale}
	
\author{Wolfgang K\"orner}
\email{wolfgang.koerner@iwm.fraunhofer.de}\affiliation{Fraunhofer Institute for Mechanics of Materials IWM, W\"ohlerstra{\ss}e
	11, 79108 Freiburg, Germany}

\author{Daniel F. Urban}
\affiliation{Fraunhofer Institute for Mechanics of Materials IWM, W\"ohlerstra{\ss}e 11, 79108 Freiburg, Germany}
\affiliation{University of Freiburg, Freiburg Materials Research Center (FMF), Stefan-Meier-Stra{\ss}e 21, 79104 Freiburg, Germany}
	
\author{Christian Els\"asser}
\affiliation{Fraunhofer Institute for Mechanics of Materials IWM,
		W\"ohlerstra{\ss}e 11, 79108 Freiburg, Germany}%
\affiliation{University of Freiburg, Freiburg Materials Research Center (FMF), Stefan-Meier-Stra{\ss}e 21, 79104 Freiburg, Germany}
	
\begin{abstract}
We present a theoretical study on the magnetic signals of structural surface defects like cracks or indents combined with inhomogeneities on the surface or subsurface inclusions
of soft ferromagnetic metals like body-centered cubic Fe or amorphous CoFeB.  We discuss limits
of early detection of small surface defects on the basis of calculated magnetic stray fields few tens of nm above the surface. The considered surface imperfections have extensions of a few nm which correspond to low multiples of the magnetic exchange lengths of Fe or CoFeB.
The detection of such small inhomogeneities requires that the sensor is about as close to the surface as the size of the inhomogeneity is.
Furthermore, the step width of a scanning sensor must be of the same size as well. Both these requirements may be fulfilled for instance by scanning microscopy with diamond nitrogen-vacancy-center quantum sensors.
\end{abstract}

\pacs{Keywords: crack detection, magnetic field measurment, NV sensor}
%
%
\maketitle

\section{Introduction}

\begin{figure}[t]
	\centering
	\setlength{\unitlength}{1mm}
	\begin{picture}(100,83)(0,0)
		\put(1,45){\includegraphics[width=8.4cm]{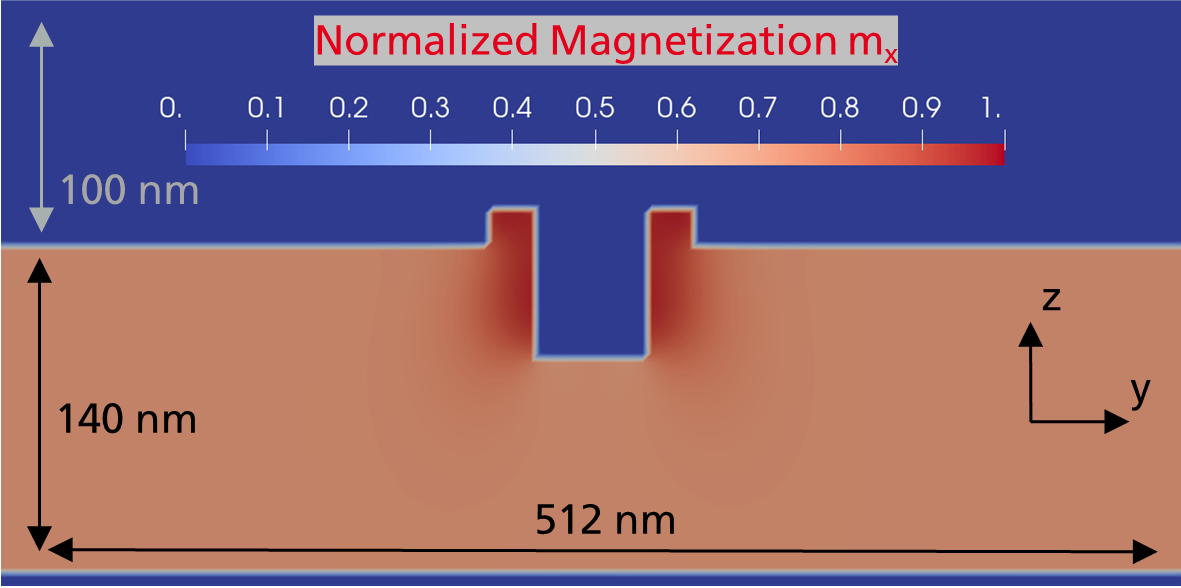}} 
		\put(1,0){\includegraphics[width=8.4cm]{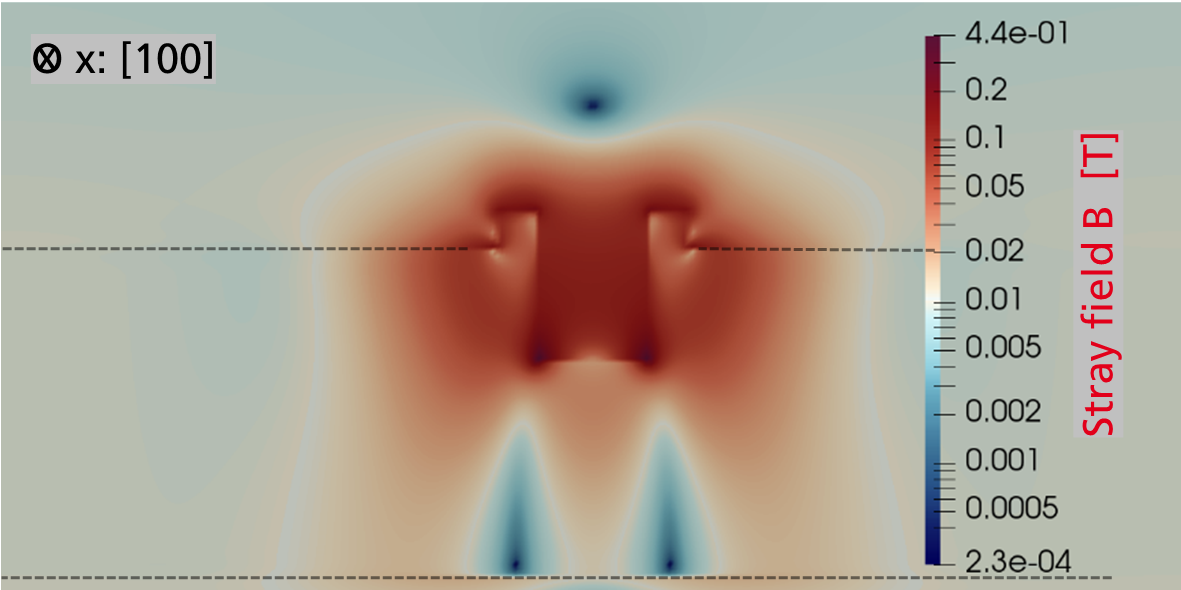}}
	\end{picture}
	\caption{Top: Distribution of the normalized magnetization m$_x$= M$_x$/$M_s$ around a cubic indent of edge length L= 48 nm with a rim. The geometry of the simulation model is described in detail in Fig. 2a. The normalized bulk magnetization points in the [110] direction, which amounts to values of approximately 1/$\sqrt{2}$ for m$_x$ and m$_y$.
		Bottom: logarithmic plot of the corresponding magnitude of the magnetic field B in Tesla. The black dashed lines indicate the 
		top and bottom plane of the magnetic metal slab in the simulation box. }\label{Fig1_example}
\end{figure}

Detection of structural surface imperfections like few nm small cracks or indents is of high technical relevance for monitoring and estimating the time to failure of structural materials. Failure of metals by fatigue or fracture often starts at the surface of a piece of metal.
Non-destructive testing (NDT) is an established approach that can reduce maintenance
costs and help to improve reliability, durability and safety of components. There are numerous macro-scale techniques like visual inspection, resistance strain gauges, piezoelectric transducers, vibrational and modal analysis, as well as
electric and magnetic sensors.\cite{esl23,kah22}

For example, at the micro-scale, cracks with an opening of only 5 $\mu$m can be detected \cite{wan00,wan02} with the
near-field microwave technique.\cite{men16}
The next step is consequently to address the nanometer scale, which is the focus of this work. To detect the origins of damage and failure as early and as locally as possible, for example at a weld seam of a steely (hydrogen) gas pipeline, or other metallic building components which are exposed to static or cyclic loading, is advantageous in various aspects. The observation and tracking of nanometer-sized defects can improve component-design or lifetime-assessment models with respect to mechanical fracture or fatigue. This is highly desirable since current codes and standards dealing with damage and failure of metals, need to be adapted to metals in heavy-duty environments \cite{rie21}, e.g. in hydrogen-gas atmospheres.\cite{ fis23, oes24}
	An actual topic is the investigation of hydrogen related local damage in microstructures of metals.   This has been studied recently in e.g.~Nickel oligocrystals by Singh et al.\cite{sin24}. The observation of initiation and propagation of microstructural damage in such macroscale specimens is of importance for improvement of understanding. However, the detection of material damage from the beginning at the nanometer scale will be even better \cite{str15, dur21, dur22, sin22}.
	With our micromagnetic model calculations we estimate magnitude and spatial distribution of magnetic stray fields due to nanometer-sized indents and further inhomogeneities on magnetic metal surfaces which is useful for the design and layout of magnetic quantum sensors for NDT of materials.

Magnetic sensors like scanning superconducting microscopy (SSM) based on SQUIDs (Superconducting QUantum Interferometer Devices), magnetic force microscopy (MFM) and scanning nitrogen-vacancy (NV) centre microscopy (SNVM)  have currently the highest spatial resolution down to 10$^{-8}$m. Furthermore, the latter three mentioned methods are sensitive to small fields down to 1 $\mu$T (for more information see e.g. the recent review of Marchiori et al.\cite{mar22} and references therein) which is sensitive enough for nm-sized defects as we will show in the following.

However, real materials have surfaces and do not contain only indents or cracks but also subsurface inclusions or surface protrutions. Thus, the magnetic stray field signal above the surface of the sample under investigation may be hard to interpret.
With our theoretical study we want to give answers to several questions which can help to optimize the techniques and to clarify the requirements for such measurements. First of all: how strong are the signals of nm-sized surface defects like cracks or indents?
Furthermore: how does the signal of an indent scale with its volume and depend on its shape? What is the influence of a rim around an indent?
What is the effect of other surface imperfections nearby?
How strong is the effect of a magnetically different subsurface inclusion below an indent?  
And finally: how does the magnetic signal depend on the axial distance between the metal surface and the sensor, and how dense does the lateral step width of the scanning sensor need to be? 

To answer these questions we have set up various micromagnetic surface models
for two prototypical soft magnetic metals, namely crystalline body-centered cubic (bcc) iron and amorphous CoFeB, which allow a systematic analysis
of features of single defects as well as combinations of these.
Already simple geometric shapes of indents can lead to complex magnetic field distributions. This can be seen in Fig.~\ref{Fig1_example} where the stray field around a cubic indent with a rim on a surface is displayed.  
Next, before we discuss Fig.~\ref{Fig1_example} and the other results of our micromagnetics simulation study in detail in section
\ref{sec_dis}, we describe
in section \ref{sec_theo} the theoretical approach and the model parameters of micromagnetics. A summary in section \ref{sec_sum} concludes the paper.

\begin{figure*}{t}
	\begin{center}
		\setlength{\unitlength}{1mm}
		\begin{picture}(220,110)(0,0)
		\put(5,0){\includegraphics[width=17cm]{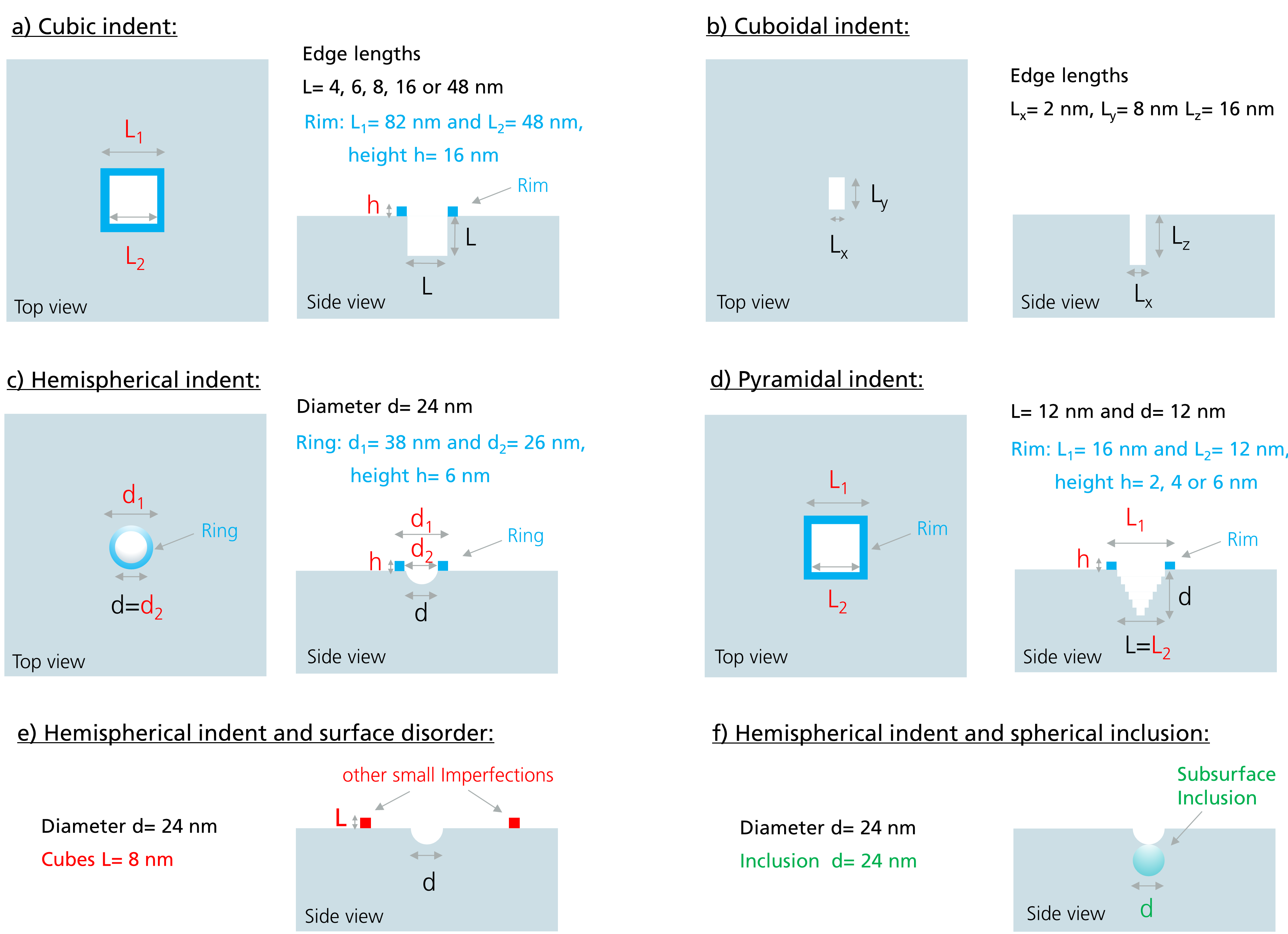}} 
		\end{picture}
	\end{center}
	\caption{(Color online) Schematic sketches of the different model systems: 
		 a) cubic indent with or without rim in the center of the simulation cell, b) cuboidal indent (with three different edge lengths), c) hemispherical indent (with or without rim), d) pyramidal indent: step size is 2 nm, resulting in a depth of 12 nm (with or without rim).
	e) hemispherical indent combined with 5 cubes with edge lengths 8 nm and randomly distributed on the surface and f) hemispherical indent with an additional spherical subsurface inclusion below the indent with different magnetic properties than the surface slab. 
		\label{Fig2_models}}
\end{figure*}

\section{Theoretical Approach}\label{sec_theo}

\subsection{Simulation Details}\label{sec_num}

For our three dimensional micromagnetics simulations we have used the software package MuMax3 which is able to treat systems with millions of discretization cells in adequate times by means of its implementation for GPUs.\cite{van14}
We solve the Landau-Lifshitz-Gilbert (LLG)\cite{lan92,gil04} equations
for the normalized magnetization $ \vec{{\rm m}}= \vec{{\rm M}}/M_s$
\begin{equation}
\frac{\partial \vec{{\rm m}}}{\partial t}=\frac{\mu_0\gamma }{1+\alpha^2}~ \vec{{\rm m}}\times \vec{{\rm H}}_{\rm eff}+ \frac{\mu_0 \gamma  \alpha}{1+\alpha^2}~ \vec{{\rm m}}\times (\vec{{\rm m}}\times \vec{{\rm H}}_{\rm eff})
\end{equation}
with the permeability of vacuum $\mu_0$, the gyromagnetic ratio $\gamma=$ 1.76086$\times 10^{11}$/(sT),  and the Gilbert damping constant $\alpha$, 
 using the RK56 (Fehlberg) solver with an adaptive time step.
In the effective field $\vec{{\rm H}}_{\rm eff}$ we take into account
demagnetization and exchange which are the dominating field contributions for
soft magnetic materials.\cite{exl21} 

The material parameters for amorphous CoFeB and bcc Fe were taken from the literature: for CoFeB the saturation magnetization is $M_s$ = 1.2$\times 10^6$A/m \cite{gu15} and the exchange constant is $A_{ex}$ = 18 pJ/m \cite{con14}, resulting in a magnetostatic exchange length\cite{mcm97} $l_{ex}=\sqrt{A_{ex}/(\mu_0M^2_{s})}\approx4.0$\ nm.
For Fe bcc the saturation magnetization is $M_s$ = 1.71$\times 10^6$A/m and 
exchange constant is $A_{ex}$ = 21 pJ/m \cite{co10} yielding an exchange length $l_{ex} \approx$ 2.4 nm.

Since the computational cost of MuMax3 is dominated by fast Fourier transform (FFT)  using the cuFFT library, which performs
best for sizes of powers of two\cite{van14} we have used discretized simulation boxes of either 512$\times$512$\times$256 nm$^3$ (large size) or 512$\times$512$\times$128 nm$^3$ (small size).   

To achieve a numerically well enough converged total energy, angles between discrete spins (normalized magnetization vectors) in adjacent cells should be less than about 30° by choosing a small enough size for the mesh cells.\cite{van14}
With a size of $2\times2\times2$ nm$^3$ for both CoFeB and Fe this angle criterion was well fulfilled.

An example of a large model system can be seen in Fig.~\ref{Fig1_example}. It consists of a three-dimensional CoFeB sheet with a thickness of $d=140$ nm in z-direction (our small model systems have $d=78$ nm), and lengths of 512 nm in both x and y directions. Above the sample surface in the simulation box we included a vacuum region of 100 nm thickness for the large systems and 50 nm for the small systems. This allows the analysis of the stray field at different heights in the following.

In order to avoid artificial boundary effects in x and y directions we used the implemented pseudo periodic boundary conditions and repeated the box 12 times in both x and y directions.\cite{van14}
Our models are thus two dimensional arrays of periodically repeated surface defect structures. Convergence tests confirmed that in order to avoid interactions between those defects their extensions have to be less than about one tenth of the x and y extensions of the simulation box.

For most of the data analysis we have used the software package Paraview (version 5.11.1).\cite{aya15}

\subsection{Model systems}\label{sec_systems}

We use a set of different model systems to investigate the influence of different aspects of surface defects on magnetic field-sensor signals. Some of them are sketched schematically in Fig. \ref{Fig2_models} and described in the following:

1. Indent on surface (only): We have modeled several indents of different shapes: cubic indents and cuboidal indents with different lengths in x, y and z directions, hemispherical indents and pyramidal indents. 

2. Indent with rim:
Indents were combined with rims of different thickness and height assuming that part of the material by indentation is redeposited along
the rim of the indent on the surface. In the case of the hemispherical indent the rim is a ring defined by an outer diameter ($d_1$), an inner diameter ($d_2$), and a height $h$.
For the pyramidal and cubic indents we have chosen quadratic rims defined by an outer edge length ($L_1$), an inner edge length ($L_2$), and a height $h$.

3. Indent with rim and additional imperfections nearby on a surface:
The inhomogeneities on the surface were modeled by arbitrarily putting five little cubes onto the flat surface.
These cubes have an edge length of 8 nm (see Fig.~\ref{Fig6_disorder}a). 

4. Indent and inhomogeneit on a surface above a subsurface inclusion:
As an additional feature we modeled subsurface inclusions by spheres below the indents with reduced saturation magnetization $M_{red}$ of 50\%, 80\% or 90\% with respect to the bulk material value $M_s$. This case is a simple model for a small precipitate (e.g., a cementite particle) or a small plastically phase-transformed region (e.g., a fcc or hcp iron particle) below an indent on a surface of e.g. the bcc iron, which has slightly different magnetic properties than iron.

\section{Results and discussion}\label{sec_dis}

The focus is on the exploration of sensing limits for the idealized case of nearly perfect surfaces of iron or CoFeB with only magnetic single-domain states.
We restrict our study to magnetic single-domain solutions because the presence of domain walls in multi-domain solutions leads to variations in the magnetic stray-field strength, which are in the range of mT, much stronger than the signals coming from nm sized indents to be studied, as described below.  
Furthermore, we restrict our analysis to low-energy states with the magnetization oriented along the surface plane. A magnetization component out of the surface plane would create a strong stray field and thus lead to unfavourable high-energy states. 

In the following we describe and discuss the results of our numerical micromagnetics simulations on how the magnetic field signals above the surface vary with sizes and shapes of indents in the surface, of indents with rims, of indents with other small particles nearby on the surface, and of indents with inclusions below them inside the surface. The simulation results for CoFeB and Fe differed in most aspects only quantitatively within 10\%. Therefore we only present the results for CoFeB  in the following.

\subsection{Signals of indents }\label{subsec_1}
For the case of cubic indents we have studied the size dependence of the magnetic field signal.
In Table \ref{tab_sig_size} the results of our simulations are summarized.
We have set the smallest cube of length $L = 4$ nm as reference for the bigger cubic indents.
For the total surface area we have taken into account the five square-shaped walls of size $L\times L$ which are created by the cubic indents.
From the results in the table we infer that the magnetic signal approximately scales rather with the volume of the indent than with the area of its walls.

\begin{table}[]
	\begin{tabular}{c c c c c}
		\hline \hline
		Edge length  ~& ~ Ratio of ~ &~ Ratio of  ~&Field signal     & Ratio of  \\
		~[nm]   ~ & wall areas  & volumes   &  [$\mu$T]  & field signals  \\
		\hline
		4  &  1    &  1     & 16.5 & 1   \\
		6  &  2.25 &   3.38  & 52   & 3.15 \\
		8  &  4    &  8     & 115  & 6.97 \\
		16 &  16   &  64    & 770  & 46.67 \\
		\hline \hline
	\end{tabular}
	\caption{Magnetic stray-field signals of cubic indents of different edge lengths L$^3$ in CoFeB in relation to their volumes and
		surface areas. The magnetic signal scales approximately with the volume of the indent.
		\label{tab_sig_size}}
\end{table}
The resulting preferred direction for the magnetization is the [110] direction in the bulk region of the surface slab model, because this arrangement minimizes the magnetic stray field pointing out of side walls of the simulation box.\cite{exl21}
If we impose a magnetization in the [100] or [010] direction the energy of the systems is higher due to the increased stray field energy originating from the magnetization pointing perpendicularly out of on one side wall of the simulation box.
As can be seen in Fig.~\ref{Fig1_example}, there are deviations from preferred [110] direction in the close vicinity of the indent (within few multiples of the exchange length).  There the magnetization is aligned preferably parallel to the surface (dark red regions correspond to [100] orientation in Fig.~\ref{Fig1_example}). 
Exl et al.\  show in Ref.\cite{exl21} how the magnetization ends up in different low-energy patterns depending on the ratio of exchange length and the edge length of a magnetic object (in their case it is a square).  Since the sizes of our investigated objects are maximally 50 nm, which is only about 25 times $l_{ex}$ of Fe and about 12 times $l_{ex}$ of CoFeB, we
always obtain magnetization arrangements that Exl et al. call leaf patterns, with preferred magnetization direction [110] and little deviations at the surfaces of the defects.
 \begin{figure}[]
	\centering
	\setlength{\unitlength}{1mm}
	\begin{picture}(100,112)(0,0)
		\put(0,110){a)} 
		\put(2,52){\includegraphics[width=8.4cm]{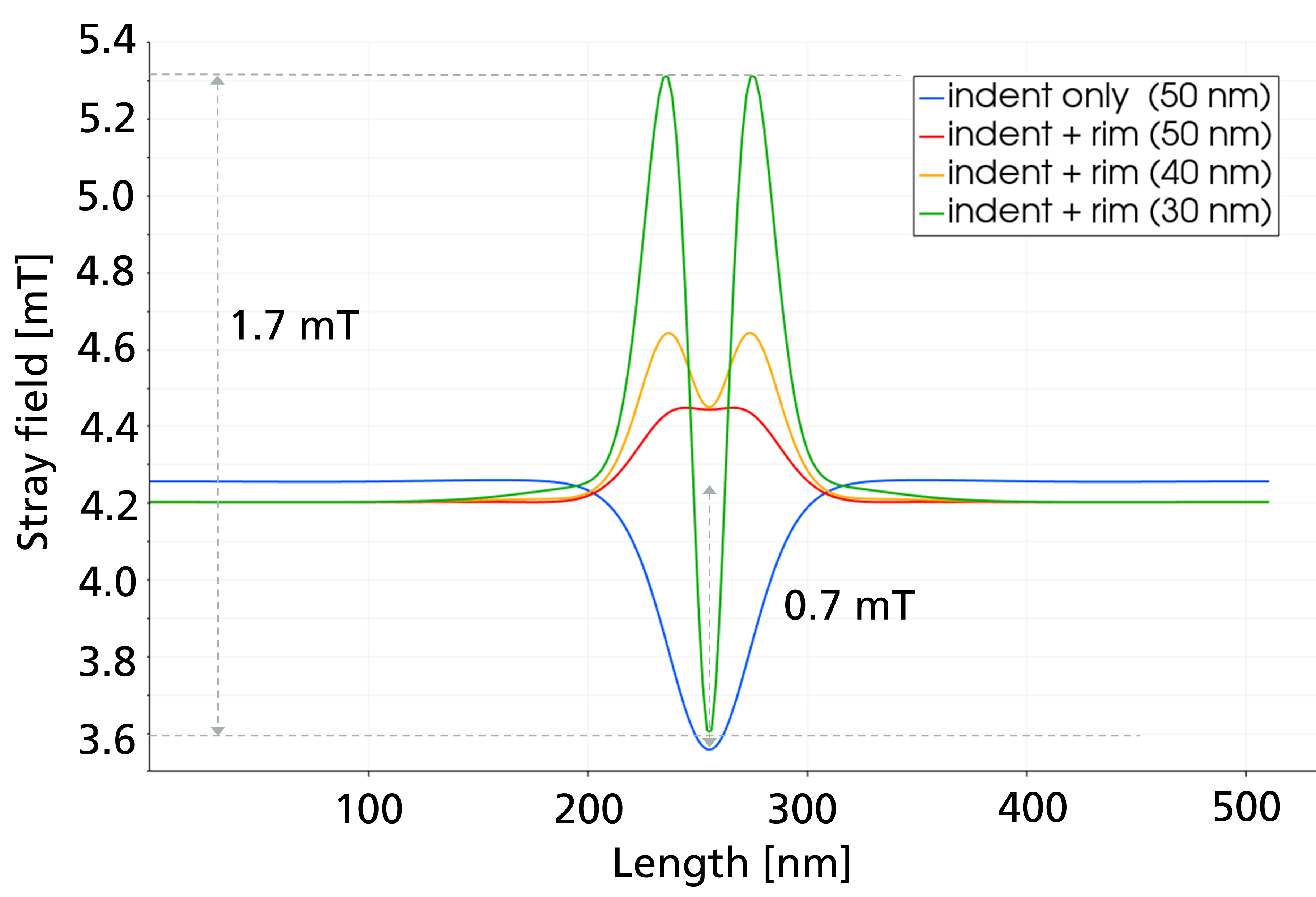}} 
		\put(0,50){b)}
		\put(3,0){\includegraphics[width=8.3cm]{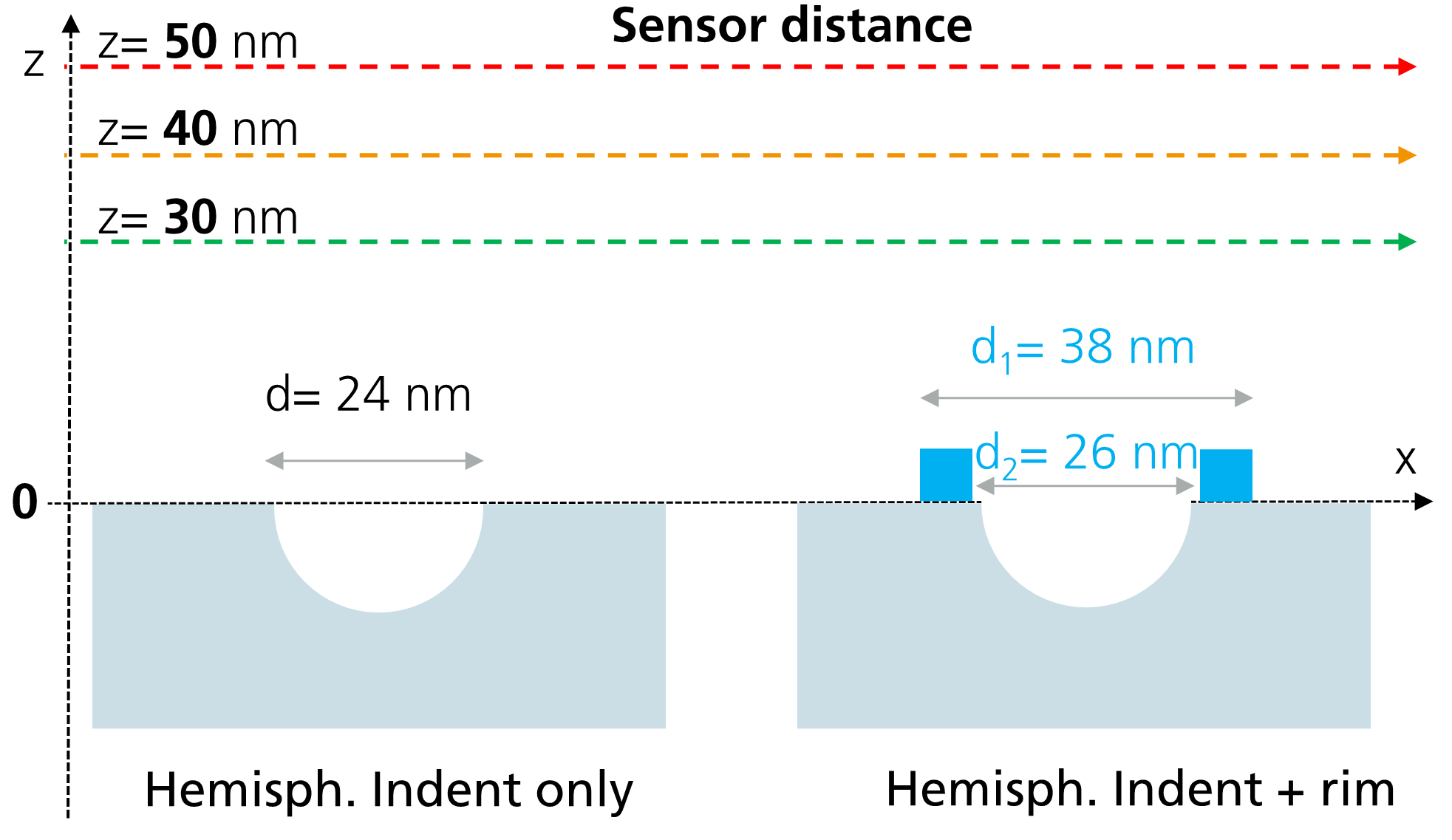}}
	\end{picture}
	\caption{a) Magnitude of the magnetic stray field B of a hemispherical indent with or without rim with the dimensions given in Fig.~\ref{Fig2_models}c) plotted for different distances above the sample. At about 40 nm height above the surface an ideal sensor starts to resolve the indent with rim (outer diameter d$_2$= 38 nm, inner diameter d$_1$=26 nm). The dip due to the indent becomes visible at 30 nm.
		b) Sketch of the arrangements of the hemispherical indents and the sensor positions: the paths along which the magnetic stray-fields are plotted in a) are indicated by the dashed arrows. }\label{Fig3_B_abs_30_40_50nm_rim_and_no_rim} 
\end{figure}

Moving from the cubic indent to cuboidal indents with different edge lengths in x and y directions, one gets magnetic field signals that differ significantly in x and y directions. The lowest-energy solution is always the one where the magnetization is pointing perpendicular to the smallest side wall of the indent: e.g. if  L$_x$ $<$ L$_y$ then A$_x$= L$_x$$\times$L$_z$ $<$
A$_y$= L$_y$$\times$L$_z$, and the magnetization points in [100] direction. 
If edge lengths of the indent are similar in all directions, like in the cases of cubic, hemispherical and pyramidal indents, then the size and the shape of the signal nearly do not depend on the direction parallel to the surface plane. A typical signal produced by a hemispherical indent of diameter d= 24 nm is plotted in Fig.~\ref{Fig3_B_abs_30_40_50nm_rim_and_no_rim}a (blue line). At 50 nm distance of the sensor above the surface one has a stray field signal of about 0.7 mT and the signal width in x or y directions is about 60 nm (we have taken the width at decay to 1/e).

In summary, we have answered the first three questions posed
	in the introduction: a nm-sized indent in a surface causes a magnetic stray-field signal in the range from few $\mu$T to few mT, the signal scales approximately with the volume of the indent, and there is a directional dependence of the field signal if the shape of the indent is strongly anisotropic like in the case of a cuboidal indent.

\subsection{Signals of indents with rims}\label{subsec_2}

\begin{figure}[]
	\centering
	\includegraphics[width=8.7cm]{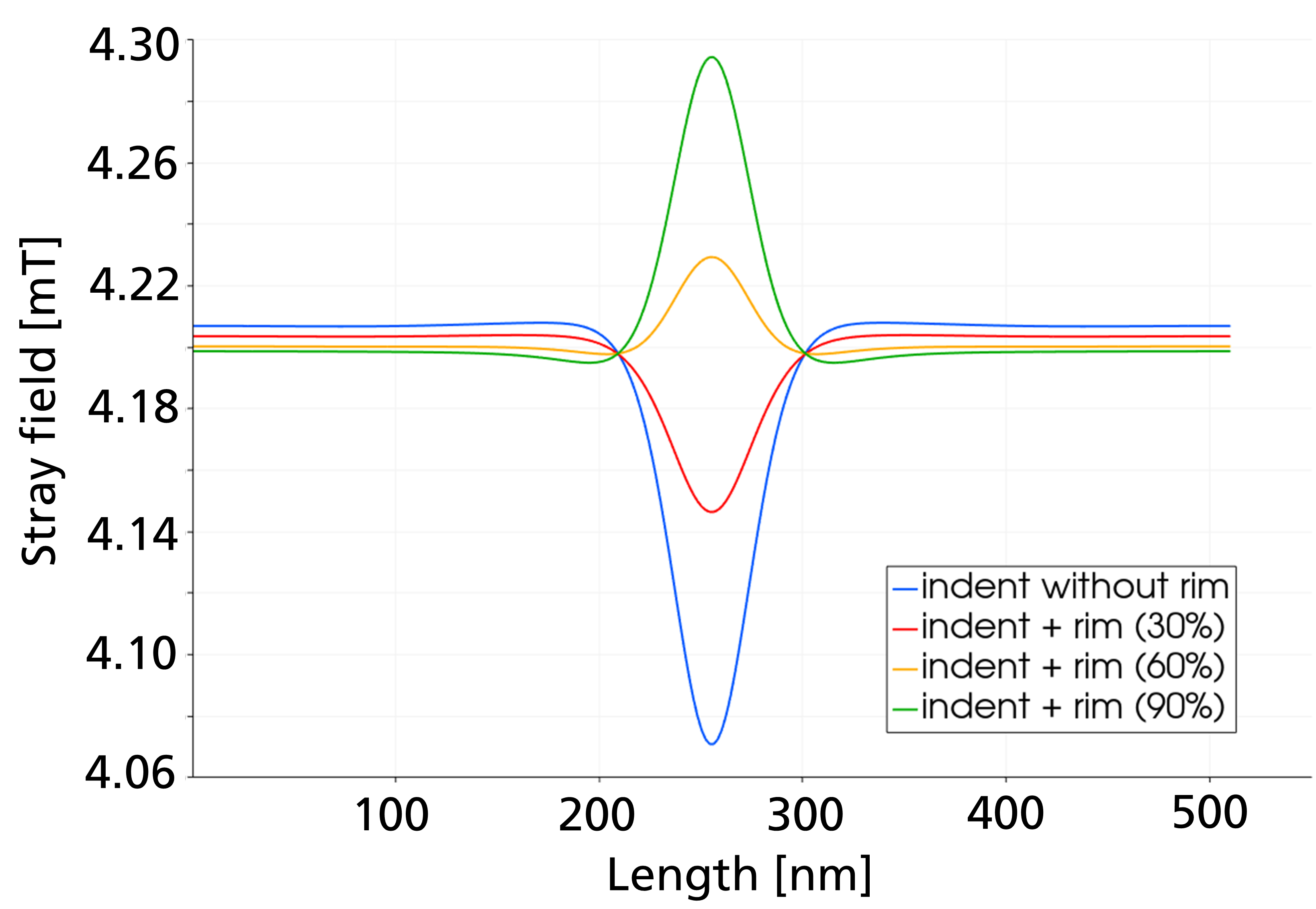}
	\caption{Magnitude of the magnetic stray field B of pyramidal indents with rims of different sizes (given by rim-to-indent volume ratios) at a distance of 50 nm above the surface.  Rims can damp (30\% volume ratio), nearly compensate (60\% volume ratio) or even invert (90\% volume ratio) the sensor signal.
	The three rims have an outer extension L$_1$ of 16 nm, an inner extension L$_2$ of 12 nm and their heights are 2, 4 and 6 nm, respectively (c.f. Fig.~\ref{Fig2_models}d). } \label{Fig4_B_abs_with_31_62_rim}
\end{figure}

Depending on the material and the way an indent is created, around its edge on the surface there may be a rim of varying width, height, shape and volume.
We have modeled such rims having volumes outside the surface of about 30\%, 60\%, and 90\% of the volume of the indent inside the surface.
Figure~\ref{Fig4_B_abs_with_31_62_rim} shows that the signal of a pyramidal indent without rim (having itself a relative volume V= 100\%, see blue line) is not only weakened by the rim (having a relative volume 30\%, red line) but nearly compensated (60\%, orange line) or even reversed (90\%, green line) when the sensor is held 50 nm above the surface. 

The indent can only be resolved as a dip in the magnetic stray-field signal by approaching the sensor closer to the surface. This is shown for
the hemispherical indent in Fig.~\ref{Fig3_B_abs_30_40_50nm_rim_and_no_rim}a.
At a distance of 40 nm between sensor and surface (orange line in Fig.~\ref{Fig3_B_abs_30_40_50nm_rim_and_no_rim}a) one resolves two peaks coming from the rim.
Approaching the sensor to 30 nm the signal is even more pronounced: the dip in the middle caused by the indent is superposed with two peaks at the left and right side due to the rim.

The stray field variation of the signal of about 0.69 mT (dip to plateau)  for the indent without rim is increased to about 1.7 mT (dip to peaks) for the indent with rim. Thus, rims do not simply shade indents but for close enough sensor position the signal variation is even enhanced by rims.

\begin{figure}[b]
	\centering
	\setlength{\unitlength}{1mm}
	\begin{picture}(100,100)(0,0)
		\put(0,95){a)} 
		\put(3,43){\includegraphics[width=8cm]{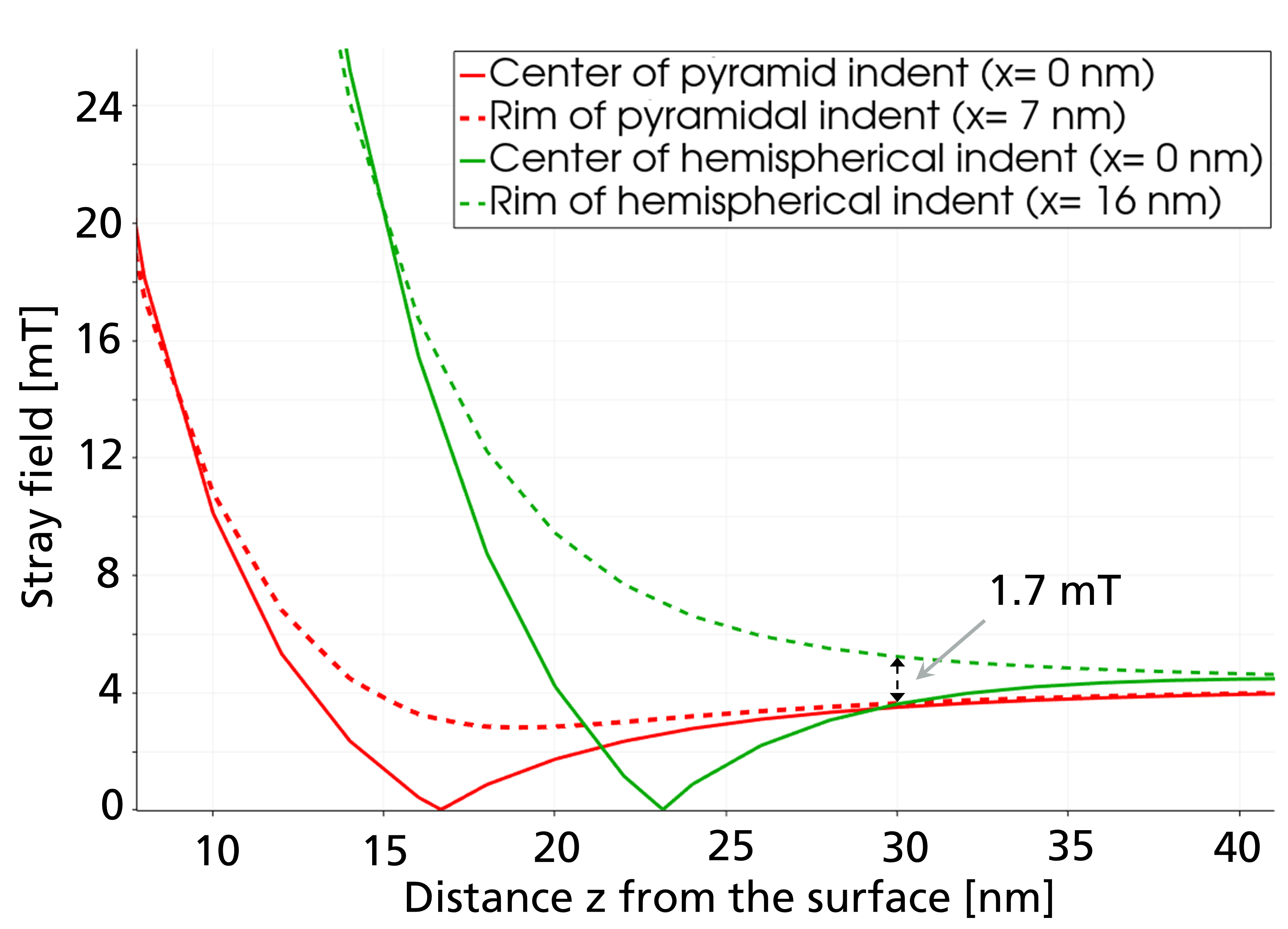}} 
		\put(0,41){b)}
		\put(2,0){\includegraphics[width=8.5cm]{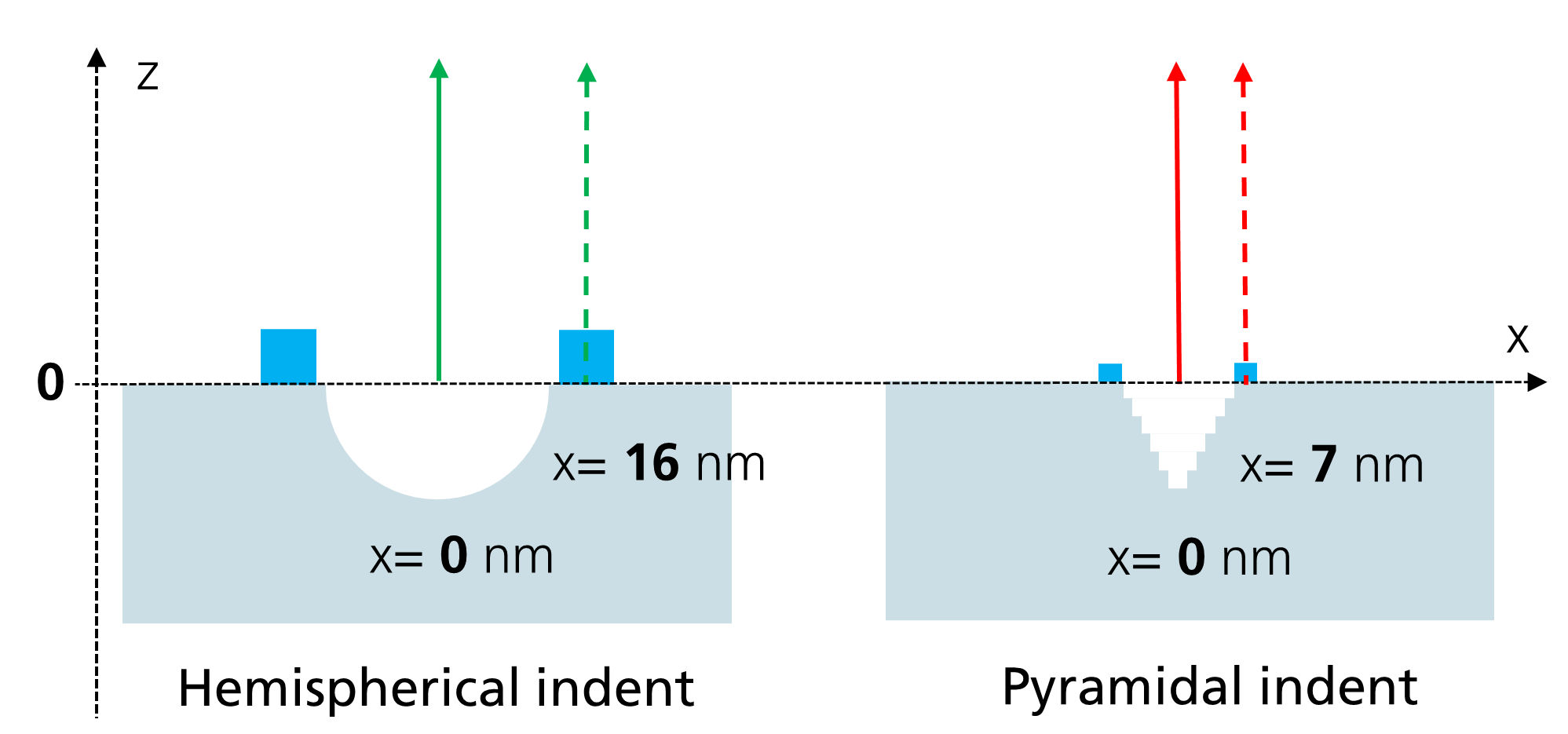}}
	\end{picture}
	\caption{ a) Dependence on the distance from the surface plane in z direction of the magnitude of the magnetic stray field B above a hemispherical indent with a large rim of 100\% volume and a pyramidal indent with a small rim of 30\% volume (see Fig. 2c and 2d for the dimensions). The solid green and the red curves show the fields perpendicularly above the centers of the indents. Moving the field sensor away from the indent centers either 7 nm (pyramid) or 16 nm (hemisphere) in x direction it is situated in the middle of the respective rim. The respective magnetic stray-field signals are plotted as dashed lines. b) Sketch of the arrangements of the indents with rims and the sensor positions: the paths along which the magnetic stray-fields are plotted in a) are indicated by the arrows. }\label{Fig5_z-dependence} 
\end{figure}

This raises the question: at what distance one reaches the maximum dip to peak variation? The answer is illustrated in  Fig.~\ref{Fig5_z-dependence} by plotting the magnitude of the magnetic stray field above the surface in z direction.
For the hemispherical and pyramidal indents we extracted the magnetic field values
directly above the center of the indents (solid vertical arrows in Fig. 5b) and directly above the middle of the rims (dashed vertical arrows in Fig 5b). The difference between the respective curves in Fig. 5a is the dip-to-peak signal variation (compare the marks 1.7 mT in Figures 3 and 5a for the hemispherical indent with rim) 
The maximum difference is always obtained where the magnitude of the magnetic stray field above the center of the indent vanishes: at $z\sim 17$ nm for the pyramidal indent and at $z\sim 23$ nm for the hemispherical indent. These distances correspond approximately to the diameters of the respective indents. 
In the plot in Fig.~\ref{Fig1_example}b this point where the magnetic stray field vanishes is situated in the center of the dark blue spot ("the head of the man")  about 54 nm above the surface.

For larger distances beween sensor and surface the difference goes to zero. Already at $z=50$ nm, about twice the diameter of the hemispherical indent, the dip and the pair of peaks cannot be resolved any more (compare the green lines in Fig 5a) with the red line in Fig.~\ref{Fig3_B_abs_30_40_50nm_rim_and_no_rim}a).   

From our calculations we derive the rule of thumb that the sensor has to be approximately at a height $z \sim d$ where $d$ is the diameter of the inhomogeneity if one wants to resolve the surface modification by indents with rims.

From the experimental perspective the positioning of a sensor to such a short distance above a metal surface is a very ambitious task.
To our knowledge the best signal resolutions were achieved by
scanning-probe-microscopy set-ups for distances between 15 and 25 nm with SNVM.\cite{cha17,ari18} 
The advantage of the SNVM over the SSM and MFM is the potentially very small sensor size since the single NV-center defect complex in diamond employed as the sensor is of an atomic size.
The SQUID’s loop and the size of the MFM sensor are typically in the range of micrometers.\cite{wer09}

 NV-center sensors, which have sufficiently long coherence times for a measuring process, can be implanted approximately 10 nm below the diamond surface\cite{ofo12} and the diamond surface of the sensor can in principle be put in contact with the surface of the soft-magnetic metal specimen, which means that a spatial sensor resolution of 10 nm is feasible.

\subsection{Effect of additional imperfections on the surface}\label{subsec_3}
An important complication of detecting  indents or cracks on a surface is a hardly avoidable surface imperfection. As a simple structural model for this we add five tiny cubes of edge length 8 nm (called particles in the following) 
on the otherwise flat surface with the indent.
We assume that these cubes are not too near to the indent, i.e. within a radius of a few nm, as this arrangement would have a
similar effect as the previously discussed rim on the magnetic signal of the indent.

\begin{figure}[]
	\centering
	\setlength{\unitlength}{1mm}
	\begin{picture}(100,120)(0,0)
		\put(0,115){a)} 
		\put(9,64){\includegraphics[width=6cm]{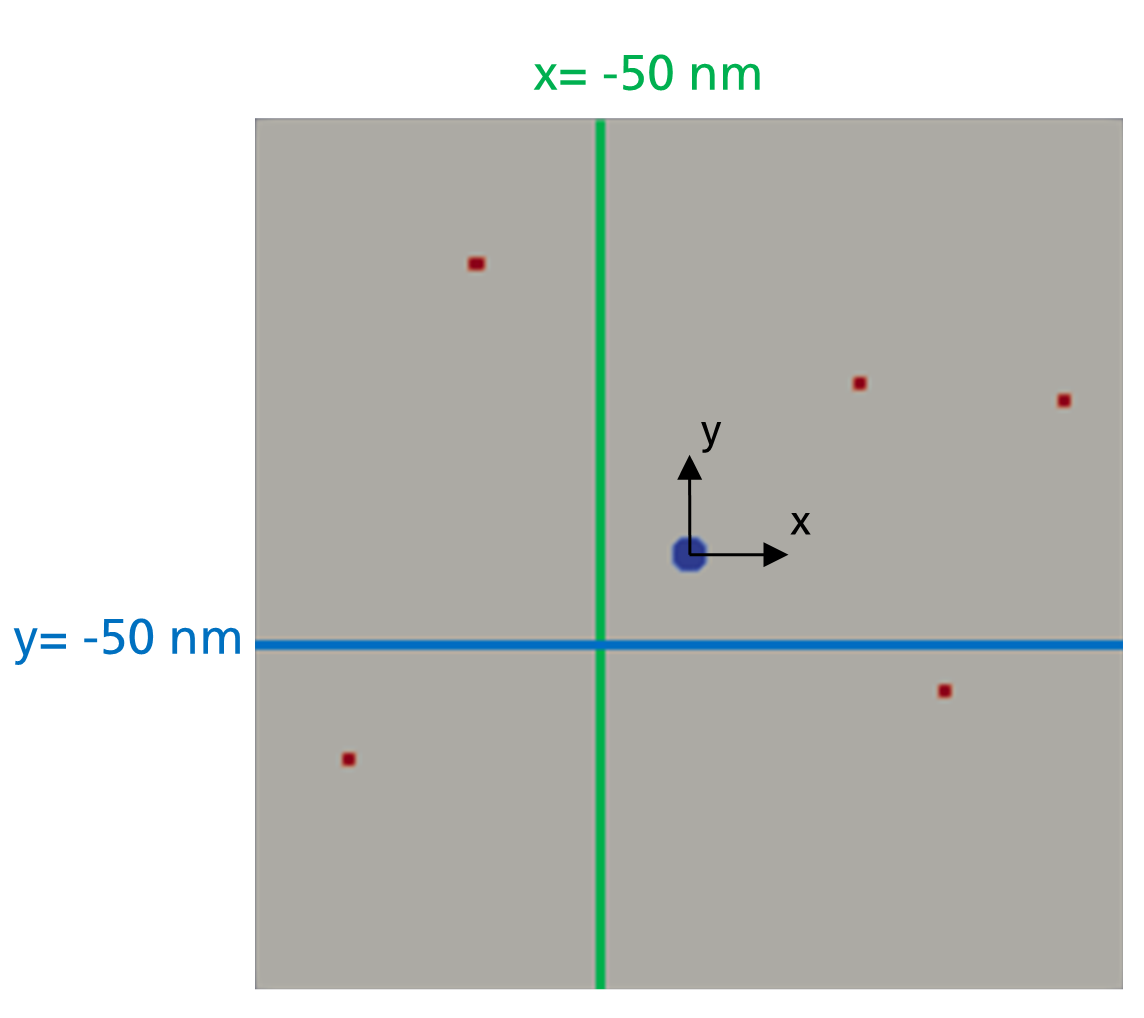}} 
		\put(0,57){b)}
		\put(-1,0){\includegraphics[width=8.7cm]{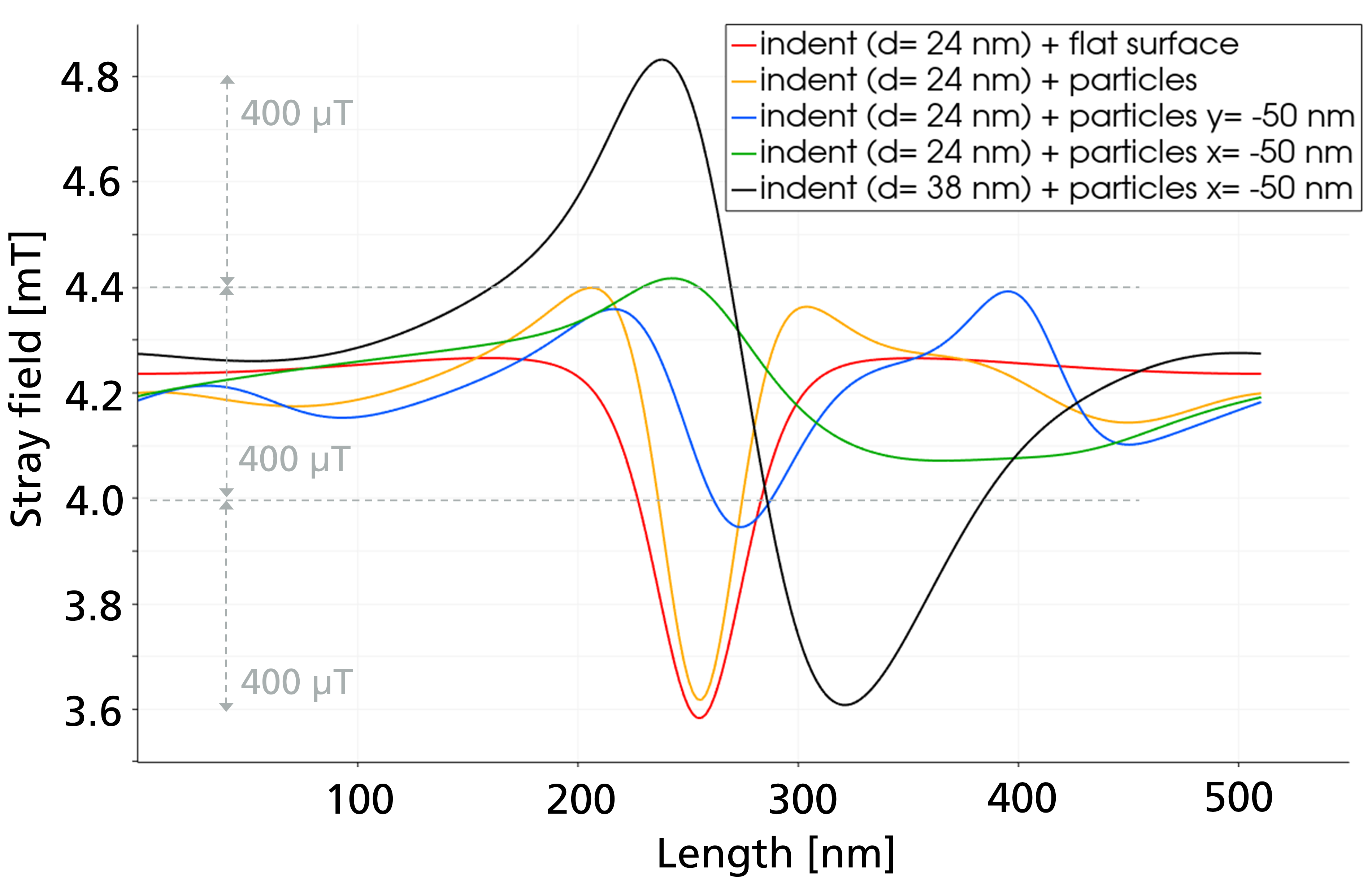}}
	\end{picture}
	\caption{a) top view on the surface: the blue spot in the centre represents the hemispherical indent and the five red spots indicate the additional, randomly set cubes (particles) on the otherwise flat surface.
		b) Magnitude of the magnetic stray field 50 nm above the surface for the hemispherical indent together with the particles on the surface: the orange and red curves are plotted
		along the x direction right across the indent. The blue, green and black curves have offsets of 50 nm as shown in the top figure a). If a sensor misses the centre of the indent by 50 nm even on a nearly smooth surface (with only few small particles nearby) the signal of the indent can not be clearly distinguished from the background variation if the indent is too small relative to the other surface objects (compare orange to green and blue curves).
		\label{Fig6_disorder} }
\end{figure}

However, as can be see from Fig.~\ref{Fig6_disorder}, even if the cubes are rather far away from the indent the signal becomes strongly disturbed and the hemispherical indent, which has an approximately seven times larger volume than the other few particles on the surface (an indent with diameter d= 24 nm), can only be seen when the sensor directly passes above the center of the indent (compare red and orange curves for flat and slightly imperfect surfaces).
By passing the indent with the sensor only 50 nm away in lateral direction leads to a too small signal which is covered by the background variation coming from the other few small particles (compare green and blue curves with orange curve).
One has to increase the size of a hemispherical indent to a diameter $d=38$ nm, which corresponds to a volume that is 28 times bigger than that of the other small particles, to get a significant stray-field signal that exceeds the background variation of about 400 $\mu$T due to the few other small particles in our model (compare green and black curves).

Hence, sensing magnetic fields above macroscopically smooth surfaces, which are microscopically inhomogeneous due to arbitrarily dispersed, nm-sized particles, require a dense scanning grid.  Since surface defects like indents or cracks can have other small surface inhomogeneities nearby, we conclude that the scanning grid needs to have a step size $d$ which is of order of the dimension  of the surface defect. This may be technically cumbersome as it results in long times needed for scanning surface areas of $\mu$m$^2$ sizes.
Furthermore, based on our simulation results we suggest that indents need to have at least about 30 times larger volume than the other objects on the surface for being reliably detectable by a scanning magnetic-field sensor of such high spatial resolution and high signal sensitivity as, e.g., a modern SNVM.\cite{cha17,ari18}

\subsection{Effect of subsurface inclusions on surface indents}\label{subsec_4}
Kryzhevich et al.~\cite{kry22} illustrated in their theoretical atomistic-simulation study on cracks in bcc iron that a small region of close-packed (hcp or fcc) iron is formed below a crack tip. Other small features with locally deviating magnetic properties may be small precipitates situated close below the surface plane with different crystal structure or chemical composition than bcc iron. Such a feature, having slightly different magnetic properties than bcc iron, can be phenomenologically incorporated in a micromagnetic simulation model by modifying the magnetic property data in a small spatial region of the simulation box.
In the following we refer to these regions with different magnetic properties as \emph{inclusions}, no matter what their origin is. 

\begin{figure}[]
	\centering
	\includegraphics[width=8.7cm]{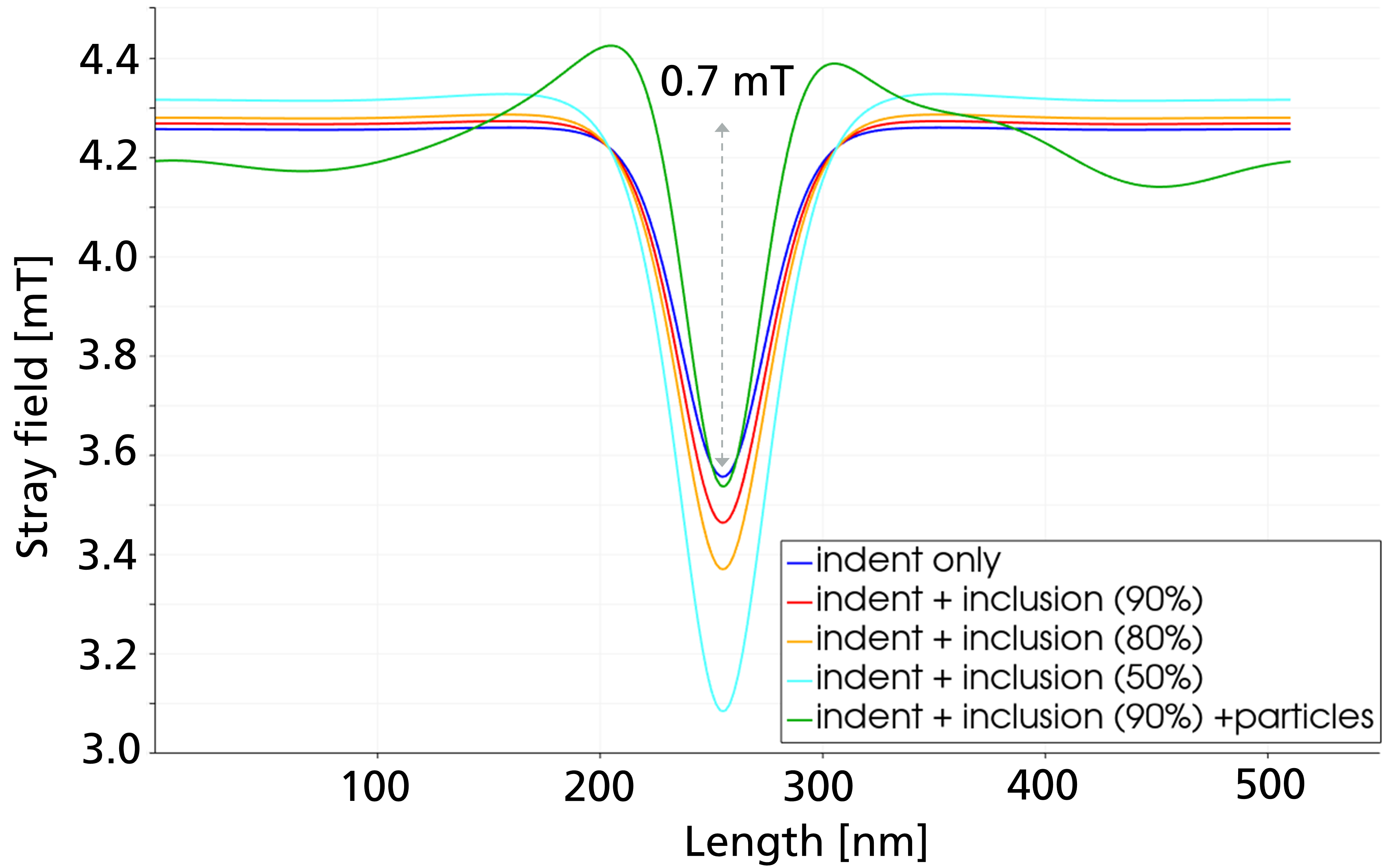}
	\caption{Magnitude of the magnetic stray field for a hemispherical indent with spherical inclusions with  90\%, 80\% or only 50\% of the bulk magnetization $M_s$ of CoFeB. For the geometric arrangement see Fig.~\ref{Fig2_models}f). \label{Fig7_indent_inclusion}}
\end{figure}

Figure~\ref{Fig7_indent_inclusion} shows how the stray-field signal dip caused by a hemispherical indent is modified by an additional spherical inclusion directly below the indent. We assume that the saturation magnetization of the inclusion is reduced either by 10\% (i.e.\ $M_{red}$ of the inclusion is 90\% of the bulk $M_s$), 20\%, or 50\% and that the exchange constant $A_{ex}$ is kept unchanged.
The dip in the signal due to the hemispherical indent is only  about 700 $\mu T$. For incremental reduction of $M_s$ by 10\% the dip becomes about 80 $\mu T$ deeper. Thus this additional change is approximately linear if the position of inclusion remains unchanged, and it is of minor importance for the total dip if $M_s$ is only modified by less than 20\%.
  
We have also simulated the magnetic stray-field signals of spherical inclusions of different volumes. The signal size of a spherical inclusion increases in good approximation linearly with its volume.
The  dependence of the signal on the depth of the inclusion is found to be not linear. It approximately follows the decay of the magnetic dipole-dipole interaction as function of the distance $r$, i.e., like $1/r^3$.

\section{Summary}\label{sec_sum}
The detection of surface defects of iron-based materials (i.e., steels), like cracks or indents with sizes of few nanometers caused for instance by mechanical fatigue or friction, is challenging since their magnetic signals are in the $\mu$T range. However, such signal sensitivities are achieved today by high-end magnetic field sensors like SSM, MFM or SNVM. 

There is a delicate balance in the magnetic signal coming from material removed below the surface plane (e.g., an indent) or added above the surface plane (e.g., a rim of an indent).  Our micromagnetic simulations illustrate that rims of indents often lead to a compensation of the sensor signal for distances between sensor and surface larger than the size of the surface defect, and clear signals can only be resolved if the sensor is close enough to the defect. As a rule of thumb, the sensor distance has to be about the same as the smallest dimension of the defect. 
Additional inclusions below indents with reduced saturation magnetization, originating for instance from a plastic deformation zone ahead of a crack tip or from a  small precipitate particle close below the surface, can lead to an enhanced stray-field signal. However, their effect remains weak if the local reduction of the magnetization is within 10\% of the saturation magnetization $M_s$ of the host metal, and if their volume of the inclusion is about the same as the volume of the indent. Furthermore, their impact is weakened due to their increased distance to the sensor (magnetic dipole interactions decay with 1/$r^3$).

A significant complication for the interpretation of magnetic stray-field signals of the surface defects of interest (indents or cracks) comes from other small and randomly dispersed imperfections on a macroscopically smooth surface. Our micromagnetic simulation results illustrate that such atomic-scale surface inhomogeneities can cover the signal of indents since their magnetic signals are decaying quickly over few tenth of nm. This means that a sensor that bypasses the center of a nm-sized indent by only about 50 nm will not give 
a clear dip or peak in the magnetic stray-field signal that is distinguishable from a background variation unless the indent is more than at least 30 times bigger in volume than other defect objects above or below the surface.
This implies the need of a dense scanning grid for the magnetic sensor to detect surface defects of few nm size reliably. However, practically this may require long measuring times with stable scanning-sensor set-ups to get magnetic stray-field maps of micrometer-sized areas of magnetic metal surfaces.

\section{Acknowledgments}

Funding for this work was provided by Fraunhofer (collaboration research project ''Quantum Magnetometry Application Laboratory'').
We thank our colleagues S. Philipp and A. Samast at Fraunhofer IWM and our project partners of Fraunhofer IAF and IPM for fruitful discussions.

\end{document}